\documentstyle[11pt,newpasp,twoside,epsf]{article}
\def\edcomment#1{\iffalse\marginpar{\raggedright\sl#1\/}\else\relax\fi}
\reversemarginpar

\begin{document}
\title{Models for infrared and submillimetre counts and backgrounds}
\author{Michael Rowan-Robinson}
\affil{Astrophysics Group, Blackett Laboratory, Imperial College of Science Technology and Medicine,
Prince Consort Road, London SW7 2BZ}

\begin{abstract}
A simple and versatile parameterized approach to the star formation history allows a 
quantitative investigation of the constraints from far infrared and submillimetre counts
and background intensity measurements.

The models include four spectral components: infrared cirrus, an M82-like starburst,
an Arp220-like starburst and an AGN dust torus.  The 60 $\mu$m luminosity function
is determined for each chosen rate of evolution using the PSCz redshift data for
15000 galaxies.  The proportions of each spectral type as a function of 60 $\mu$m
luminosity are chosen for consistency with IRAS and SCUBA colour-luminosity relations, and
with the fraction of AGN as a function of luminosity found in 12 $\mu$m samples. 
 
A good fit to the observed counts at 0.44, 2.2, 15, 60, 90, 175 and 850 $\mu$m can be found 
with pure luminosity evolution in all 3 cosmological
models investigated: $\Omega_o$ = 1, $\Omega_o$ = 0.3 ($\Lambda$ = 0), and $\Omega_o$ = 0.3, 
$\Lambda$ = 0.7.
All 3 models also give an acceptable fit to the integrated background spectrum.
The total mass-density of stars generated is consistent with that observed, in all
3 cosmological models.
\end{abstract}

\section{Introduction}
Surveys at 850 $\mu$m (Hughes et al 1998, Eales et al 1998, Barger et al 1999, Blain et al 2000,
Fox et al 2000) and the detection of the submillimetre background (Puget et al 1996, Fixsen et al 
1998, Hauser et al 1998) have opened up high-redshift dusty star-forming galaxies to view.
Here I review what source counts and the background radiation 
at infrared and submm wavelengths can tell us about 
the star formation history of the universe.  Madau et al (1996) showed how the ultraviolet
surveys of Lilly et al (1996) could be combined with information on uv dropout
galaxies in the Hubble Deep Field to give an estimate of the star formation history
from z = 0-4.
However these studies ignored what is already well-known from
far infared wavelengths, that dust plays a major role in redistributing the
energy radiated at visible and uv wavelengths to the far infrared.

Subsequent to the Madau et al analysis several groups of authors have argued that the 
role of dust is crucial in estimates of the star formation rates at high redshifts.
Rowan-Robinson et al (1997) derived a surprisingly high rate of star formation at z = 0.5-1 
from an ISO survey of the Hubble Deep Field (HDF). Subsequent ISO estimates by
Flores et al (1999) confirmed the need to correct for the effects of dust in estimates of 
star-formation rates.  Large extinction correction factors (5-10) 
have also been derived at z = 2-5 by Meurer et al (1997, 1999), Pettini et al 
(1998) and Steidel et al (1999).

Here I report the results of a parameterized approach to the star formation
history of the universe, which allows a large category of possible histories to be
explored and quantified. The parametrized models can be compared with a wide range
of source-count and background data at far infrared and submillimtre wavelengths to
narrow down the parameter space that the star formation history can occupy.  The approach
is similar to that of Blain et al (1998) and Guiderdoni et al (1998), 
but differs in key respects outlined below. The models of
Franceschini et al (1997) invoke a new population of heavily obscured high redshift
starbursts, designed to account for the formation of spiral bulges and ellipticals, whereas
I am testing whether the submillimetre background and counts can be understood in terms of a single
population of evolving star-forming galaxies. A full account is given by Rowan-Robinson (2000,
ApJ subm.). 

A Hubble constant of 100 $km/s/Mpc$ is used throughout.

\begin{figure}
\plottwo{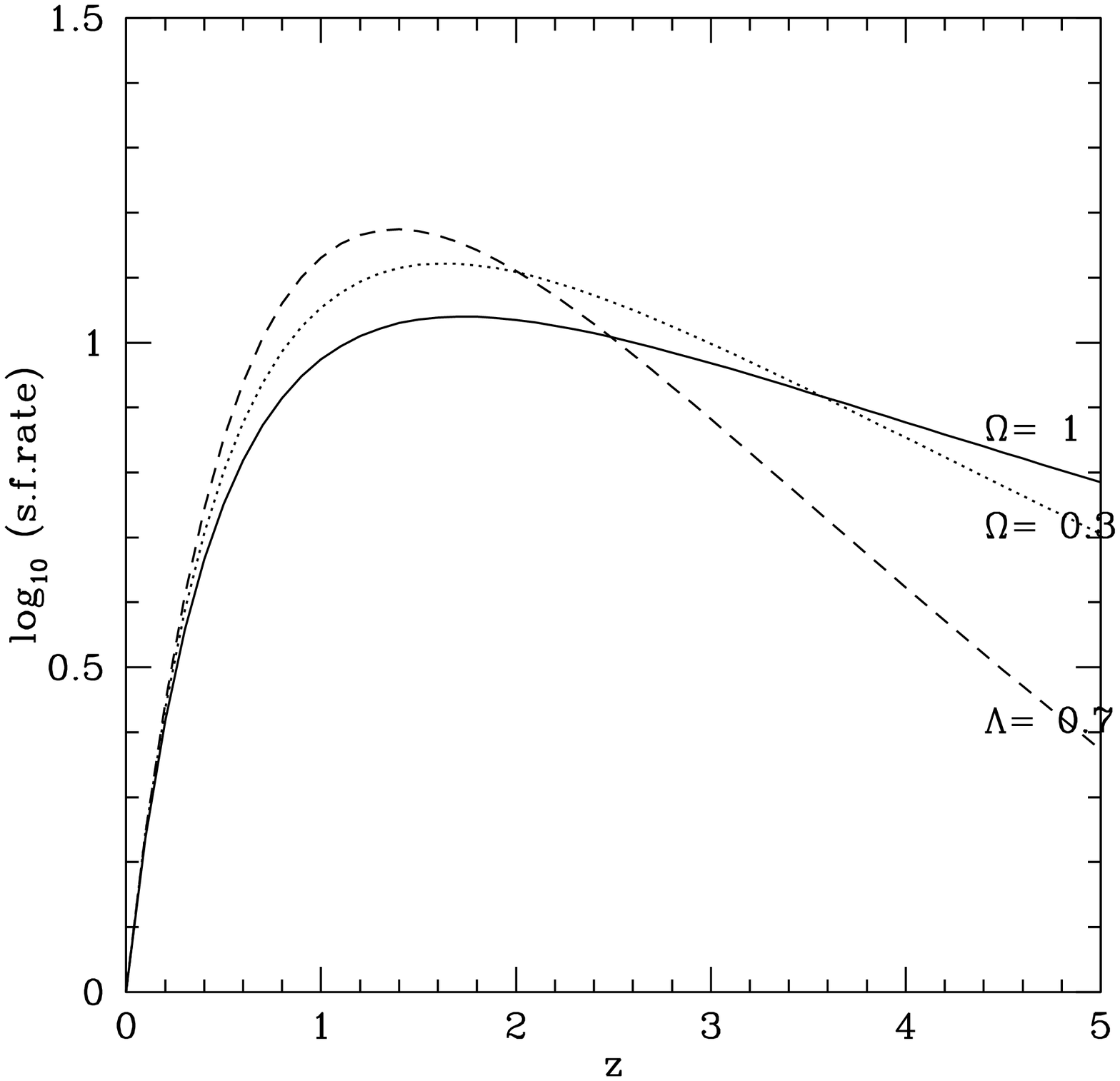}{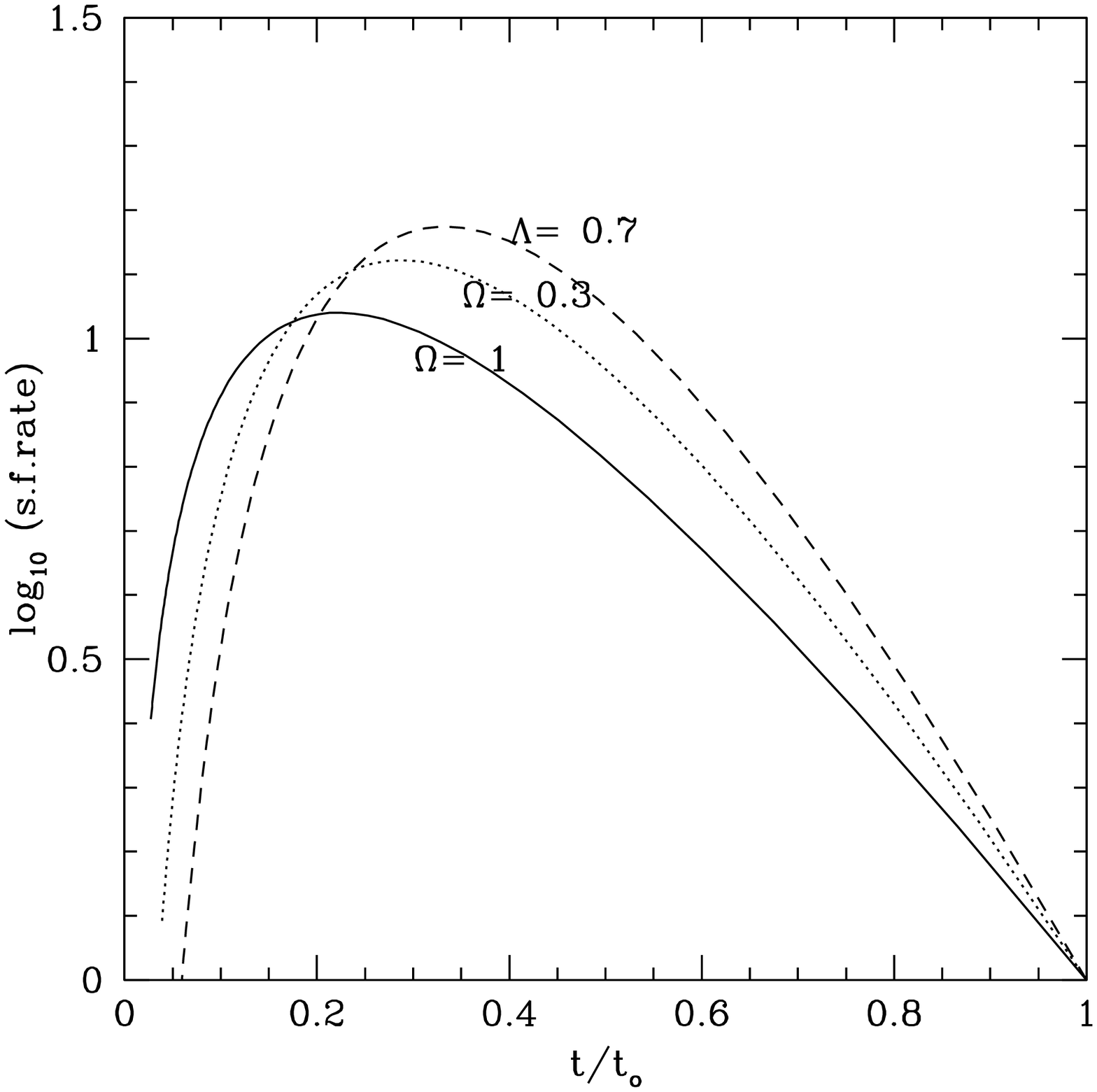}
\caption{
(L) Star formation histories of the form eqn (1), as a function of z, from best-fit models for 
far infrared and submm counts and background for
$\Omega_o = 1$, (solid curve, (P,Q) = (1.2,5.4)); $\Omega_o = 0.3$, (dotted
curve, (P,Q) = 2.1, 7.3)); $\Lambda = 0.7$, (broken curve, (P,Q) = (3.0, 9.0)).
(R) Same, but as a function of cosmic time, t.  }
\end{figure}

\begin{figure}
\plottwo{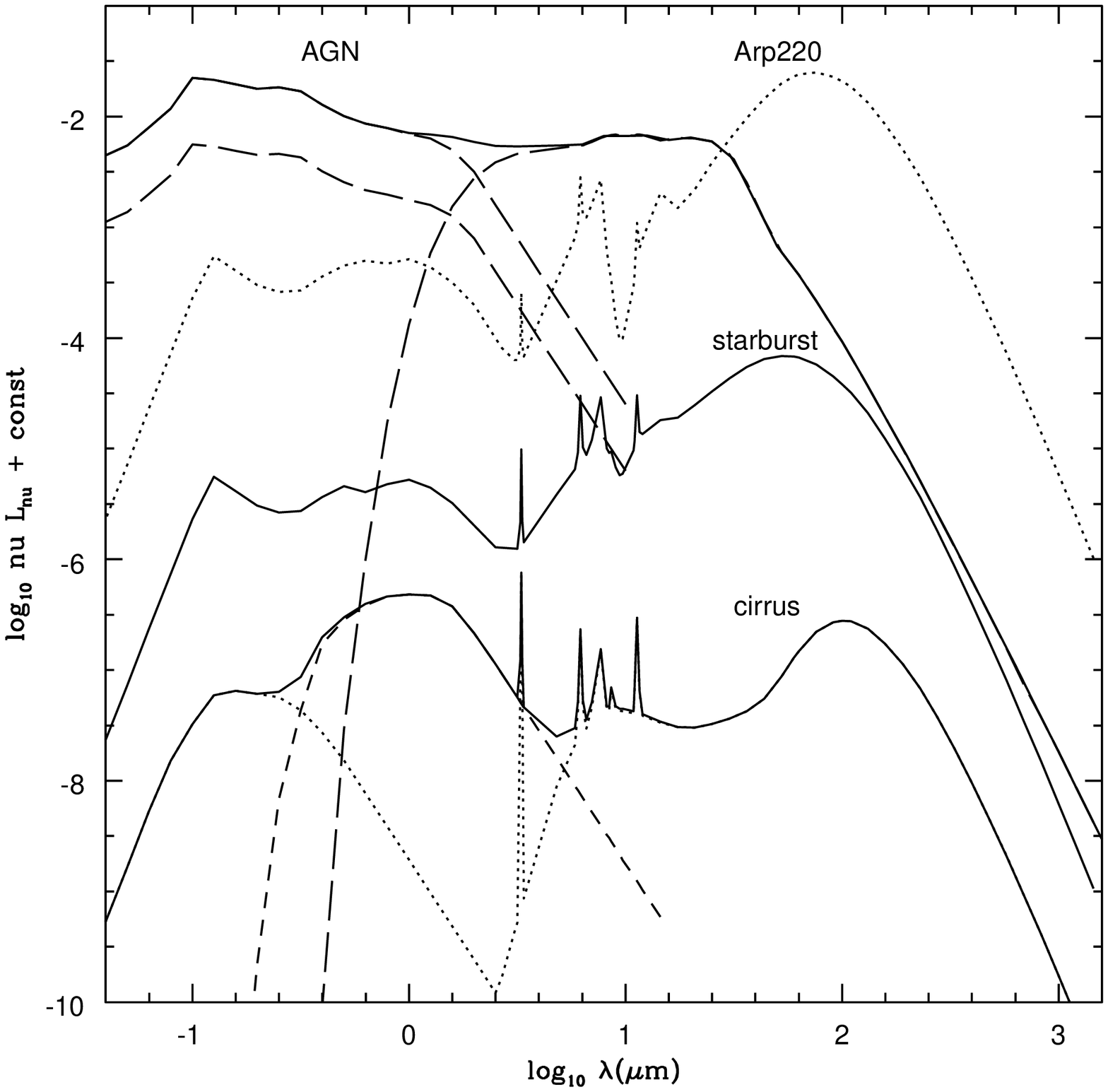}{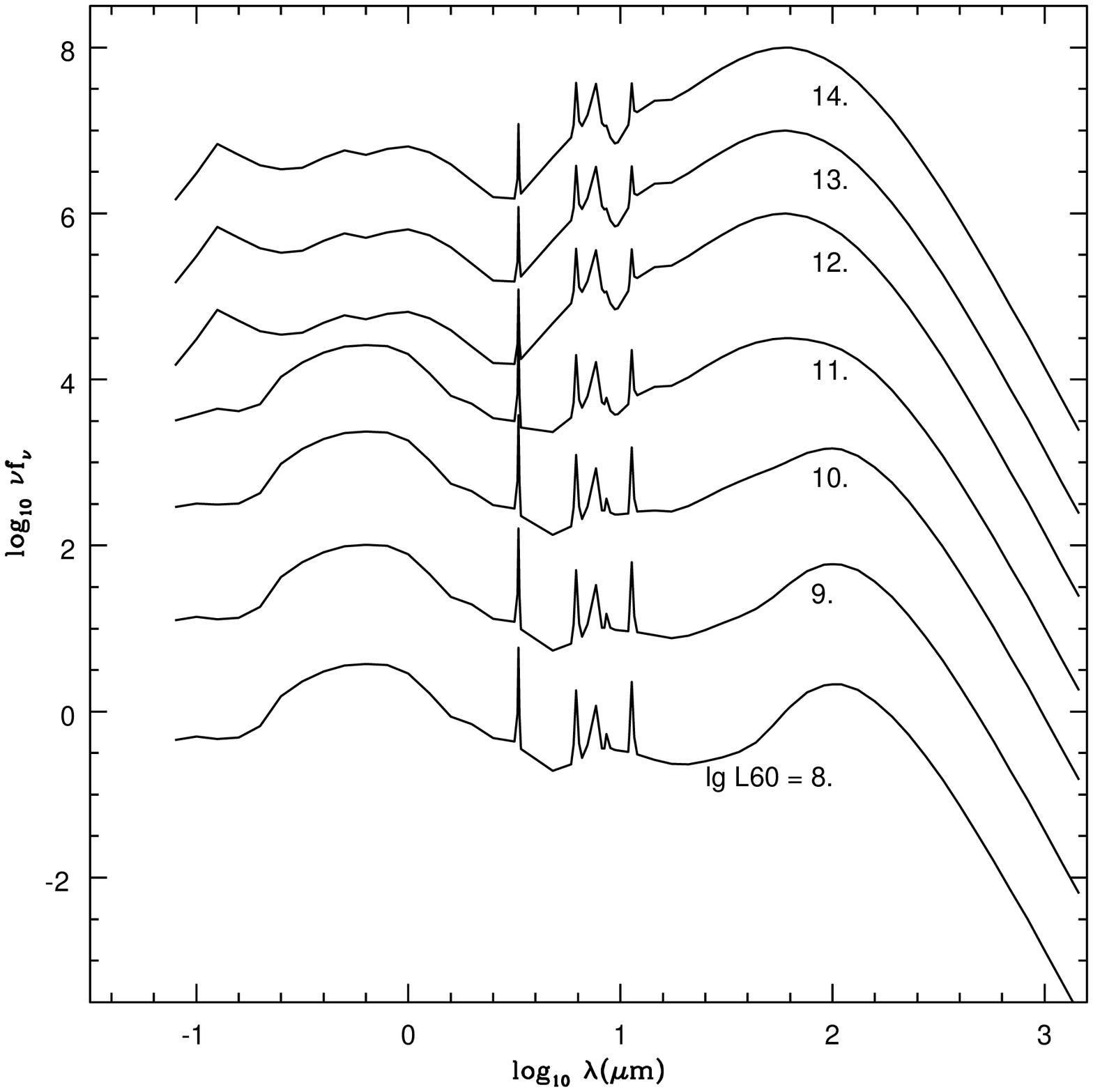}
\caption{
(L) Adopted spectral energy distributions for the four components adopted in this study: 
cirrus (with optical emission split into low-mass (broken curve) and high-mass
(dotted curve) stars), M82-starburst, A220-starburst (models from Efstathiou et al 2000), 
AGN dust torus (model from Rowan-Robinson 1995), showing assumed optical/ir ratio at
log L60 = 14 (upper curve) and 8.
(R) Average sed as a function of 60 $\mu$m luminosity, ranging from
$log_{10} (L_{60}/L_{\odot})$ = 8 to 14.}
\end{figure}

\section{Parametrized approach to star formation history}

In this paper I present a parameterized approach to the problem, investigating a wide
range of possible star formation histories for consistency with counts and background
data from the ultraviolet to the submm.

The constraints we have on the star formation rate, $\dot{\phi_{*}}$(t) are that
\medskip

(i) it is zero for t = 0

(ii) it is finite at $t = t_{o}$

(iii) it increases with z out to at least z = 1 ( and from (i) must eventually decrease
at high z).
\medskip

A simple mathematical form consistent with these constraints is

\medskip

$\dot{\phi_{*}}(t)/\dot{\phi_{*}}(t_{o})$ = exp Q(1-$(t/t_{o}))$ $(t/t_{o})^P$  (1)

\medskip

where P and Q are parameters 

( P $>$ 0 to satisfy (i), Q $>$ 0 to satisfy (iii)).  I assume that $\dot{\phi_{*}}(t)$ = 0
for z $>$ 10.

Equation (1) provides a simple but versatile parameterization of the star formation history, 
capable of reproducing most physically realistic, single-population scenarios.  
Figure 1 shows models of type eqn (1)
derived from fits to far infrared and submillimetre source-counts and
background intensity in different cosmological models (see below).  Models of
the form (1) will not, however, be able to reproduce the very sharply peaked scenario ('ED') of
Dwek et al (1998), or the two-population model of Franceschini et al (1997).  In the scenarios
modelled by eqn (1) the formation of bulges and ellipticals are part of an evolving spectrum of 
star-formation rates. 

The physical meaning of the parameters is as follows.  The exponential term describes 
an exponential decay in the star formation rate on a time-scale $t_{o}/Q$, which can
be interpreted as a product of the process of exhaustion of available gas for star formation
(a competition between formation into stars and return of gas to the interstellar
medium from stars) and of the declining rate of galaxy interactions and mergers at later
epochs.  This parameter is the same as that used in the galaxy sed models of Bruzual and Charlot (1993).
The power-law term represents the build-up of the rate of star
formation due to mergers of fragments which are incorporated into galaxies.  P measures
how steeply this process occurs with time.  The ratio P/Q determines the location of the peak in the
star formation rate, $t_{peak}$, since $t_{peak}/t_o = P/Q$.  

A very important assumption in the present work is that the star formation rate should vary
smoothly with epoch.  Several earlier assumptions have assumed, for mathematical
convenience, discontinuous changes of slope in the star formation rate (eg Pearson and
Rowan-Robinson 1996, the Blain et al 1998 'anvil' models).  Such discontinuities are
highly unphysical and I have eliminated them in this work.  

We might expect that the cosmological model could have a significant effect on the
relationship between predicted counts and predicted background intensity, since
the latter is sensitive to how the volume element and look-back time change with
redshift.  To test this I have explored models with (a) $\Lambda$ = 0, for which all the required 
formulae are analytic (specifically, here, models with $\Omega_o$ = 1, 0.3), and (b) k = 0, 
for which some of them are (specifically, $\Lambda$ = 0.7, $\Omega_o$ = 0.3). 
  
\section{60 $\mu$m luminosity function and evolution derived from IRAS PSCz survey}

Given an assumed (P,Q) I then determine the 60 $\mu$m luminosity function, using the
IRAS PSCz sample (Saunders et al 1999).  I fit this with the form assumed by Saunders et al (1990)

\medskip
$\eta(L) = C_{*} (L/L_{*})^{1-\alpha} e^{-0.5[log_{10}(1 + L/L_{*})/\sigma]^{2}}$ .  (2)

It is not clear that any previous studies have correctly taken account of the need to change
the 60 $\mu$m luminosity function as the rate of evolution is varied.  The study of
Guiderdoni et al (1998) explicitly violates the known constraints on the 60 $\mu$m
luminosity function at the high luminosity end and as a result the models predict far too many
high redshift galaxies at a flux-limit of 0.2 Jy at 60 $\mu$m, where substantial redshift surveys have
already taken place.  Blain et al (1998) state that they have determined the luminosity
function from consistency with the 60 $\mu$m counts, but this process does not automatically
give detailed consistency with existing redshift surveys.

We can also use the PSCz data to determine a range of consistency for (P,Q), using the
$V/V_m$ test.  The predicted uncertainty in $<V/V_m>$ for a population of n galaxies,
$(12n)^{-1/2}$ can be used to assign a goodness of fit for each set of (P,Q) values.

\section{Assumed infrared and submm spectral energy distributions}

To transform this 60 $\mu$m luminosity function to other wavelengths, we have to
make an assumption about the spectral energy distributions.  I have explored a variety
of assumtions about these seds: (a) a single sed at all luminosities representative 
of starbursts; (b) composite seds which are a mixture of four components, 
a starburst component, a 'cirrus'
component, an Arp 220-like starburst, and an AGN dust torus (Rowan-Robinson 1995), with the proportions of
each depending on 60 $\mu$m luminosity.  Neither of these approaches gave 
satisfactory results and it was not possible to find a simultaneous fit to all the far infrared
and submillimtre counts and background spectrum in any cosmological model.
Finally I have derived counts and background spectrum separately for each of the 
four components and then summed.  This approach finally gave satisfactory fits to all
available data.  It also allows a correct determination of the redshift distribution
of each type separately as a function of wavelength and flux-density, and the
proportion of each type contribution to the counts at any flux-density or to the background.  

I have used the latest predictions
for infrared seds of these components by Efstathiou et al (2000), and added
near ir-optical-uv seds corresponding to an Sab galaxy for the cirrus and an HII galaxy
for the starburst component, respectively.  The starburst model is a good fit 
to multiwavelength data for M82 and NGC6092 (Efstathiou et al 2000), and also to far infrared
and submillimetre data for luminous starbursts (Rigopoulou et al 1996).

The normalization
between far infrared and optical-uv components is determined by  L(60$\mu$m)/L(0.8$\mu$m) = 0.15
for the cirrus component, 5.3 for the starburst component, and 49 for the Arp 200
component.  For the AGN component I assume that L(10$\mu$m)/L(0.1($\mu$m) = 0.3 (cf
Rowan-Robinson 1995) for the most luminous AGN, and that this ratio increases with decreasing 
luminosity to account for the fact that the mean covering factor is higher at lower
luminosities.  So lg {L(10$\mu$m)/L(0.1($\mu$m)} = -0.52 + 0.1*(14.0-log L60).

For the cirrus component I have, somewhat arbitrarily, divided the optical
sed into a contribution of young, high-mass stars ($\lambda \leq 0.4 \mu m$) and a
contribution of old, low-mass stars ($\lambda \geq 0.4 \mu m$) (see Fig 2).  The former are assumed 
to trace the
star formation rate, but the latter trace the cumulative star formation up to the epoch
observed.  This treatment, though approximate, allows a reasonable prediction of the K- 
and B-band counts.  The two components in Fig 2 can be modelled, assuming $L \propto M^3$,
blackbody seds with $T \propto L^{1/2}$, and with a Salpeter mass function,
with mass-range 0.1-1 $M_{\odot}$ for the low-mass star component, and 8-40 $M_{\odot}$
for the high-mass star component.      

The proportions of the four components (at 60 $\mu$m) as a function of luminosity, $t_i(L_{60})$, 
have been chosen
to give the correct mean relations in the S(25)/S12), S(60)/S(25), S(100)/S(60) and S(60)/S(850)
versus L(60) diagrams.  Where predictions are being compared with IRAS 12 $\mu$m
data, or (later) with ISO 15 $\mu$m or 6.7 $\mu$ counts, account is taken of the width of
the relevant observation bands by filtering with a top-hat filter of appropriate half-width.
(0.23, 0.26 and 0.16 at 6.7, 12, and 15 $\mu$m respectively).  Otherwise observations were assumed to be
monochromatic.
The relative proportion of the 60 $\mu$m emission 
due to AGN dust tori as a function of L(60) is derived from the luminosity functions
given by Rush et al (1993) (see Fig 4).
The resulting mean seds as a function of L(60) are shown
in Fig 2.  Luminosity functions at different wavelengths are shown in Figs 3-4.  There is good 
agreement with measured luminosity
functions at wavelengths from 0.44-850 $\mu$m.  
  The fit to the 850 $\mu$m luminsity function of Dunne et al (2000) is 
impressive, since the transformation from 60 to 850 $\mu$m is based only on choosing the
$t_i(L_{60})$ to give the correct average S(100)/S(60) as a function of $\L_{60}$.  
  It is also impressive that luminosity functions
derived in the far infrared can fit the data at 0.44 $\mu$m (B band): the only freedom in the models
to fit the B-band luminosity functions and counts is the amplitude of the optical sed relative
to the far infrared.  

Clearly it will be
important to have submillimetre data for a wide range of normal and active galaxies to
test and improve these seds.  But the approach of using accurate radiative transfer models, 
with realistic assumptions about dust grains,
which have been verified with observations of known galaxies, seems superior to 
modelling the sed as a blackbody with power-law dust grain opacity in which
the dust temperature is treated as a free parameter (as in Blain et al 1998).
The latter approximation can only be valid for rest-frame wavelengths greater than 60 $\mu$m, ie for
accurate prediction of counts and background intensities at wavelengths $>$ 200 $\mu$m.
Useful predictions can certainly not be made at 15 $\mu$m without explicit treatment of PAHs.
These criticisms do not apply to the studies of Guiderdoni et al (1998), Dwek et al (1998),
Xu et al (1998), Dole et al (2000), 
whose assumed seds are similar to those used here.  

I have also assumed that the same luminosity evolution function should be applied to the whole
60 $\mu$m function, ie to AGN, 'cirrus' and 'starburst' components.  I have investigated
the effect of making the 
switchover in proportions of different types of component at a fixed luminosity,
so that there is in effect a strong increase in the proportion of galaxies that are starbursts  
(or contain AGN) with redshift.  However this did not permit a fit to all the available
data.

A substantial part of
the illumination of the cirrus component in spiral galaxies is due to recently formed
massive stars, part of whose light escapes directly from star-forming regions despite the
high average optical depth in these regions.  In the starburst models of Efstathiou et al (2000), 
this corresponds to the late stages of their starburst models.  If the typical starburst luminosity was
greater in the past then the emission from interstellar dust in the galaxy would also be
correspondingly greater.  I have not at this stage considered the evolution of the seds 
of each component with redshift (see discussion in section 8).    

It is possible that the evolution of AGN differs from that
of starburst at z $>$ 3, but this will have little effect on the far infrared counts and
background (there could be a significant effect at 15 $\mu$m, which will be worth further study).

Elliptical galaxies are not treated explicitly, though their star formation rate must have been
much greater in the past than at present.  I am assuming that ellipticals are quiescent
starburst galaxies, that their star formation proceeded in much the same way as we
see in current live starbursts, and that their star formation episodes are part of the
evolution history quantified here.  We have to think of this history as a series of short-
lived fireworks taking place in different galaxies at different times.  Similarly this 
approach does not track the different spiral types separately, but only in a global average
at each epoch.

\section{Combined fits to 60, 175 and 850 $\mu$m counts, and 140-750 $\mu$m background, 
and determination of P,Q}

I can now predict the counts and background intensity at any wavelength and by
comparing with observed values, constrain loci in the P-Q plane.  To determine
(P,Q) for any given cosmological model, I combine the constraints found at 60 $\mu$m
from the PSCz (section 4 above) with constraints from (1) deep counts at 60 $\mu$m (50 mJy),
(2) the observed source-counts at 850 $\mu$m at 1 and 4 mJy, (3) the background
intensity at 140, 350 and 750 $\mu$m.  For all 3 cosmological models,
values of (P,Q) can be found which give a satisfactory fit to all the available data.
As emphasized above, this outcome does depend strongly on the assumptions made about the seds.

An important constraint on the models is that the total mass of stars produced in galaxies
should be consistent with the mass of stars observed, $\Omega_{*} = 0.003 \pm 0.0009 
h^{-1}$ 
(Lanzetta et al 1996), and that it should be less than the total density of baryons
in the universe, $\Omega_* \le 0.0125 \pm 0.0025 h^{-2}$ (Walker et al 1991).  

We can calculate the total mass-density of stars from the 60 $\mu$m luminsity density using
eqn (7) of Rowan-Robinson et al (1997), modified to take account of the latest Bruzual 
and Charlot galaxy evolution models (Rowan-Robinson 2000), from

\medskip
$\Omega_* = 10^{-11.13} \xi h^{-2} l_{60} (t_o/10 Gyr)$  (5)

\medskip
where $l_{60}$ is the luminosity density in solar luminosities per $Mpc^3$,

$\xi = \int_0^1 [\dot{\phi_*(t)}/\dot{\phi_*(t_o)} ] d(t/t_o)$,

and the assumed fraction of opt-uv light being radiated in the far infrared has been assumed to be 
$\epsilon = 2/3$.

The models which fit the counts and background for $\Omega_o$ = 1 (P=1.2, Q=5.4), 
$\Omega_o$ = 0.3 (P=2.1, Q=7.3), and $\Lambda$ = 0.7 (P=3.0, Q=9.0)
give $l_{60}$ = 4.3, 4.4 and 4.1x$10^7 h L_{\odot} Mpc^{-3}$
respectively, and $\xi$ = 5.70, 6.66, 7.25, so the corresponding values of
$\Omega_*$ are 0.0027, 0.0032 and 0.0033 $h^{-1}$, for $t_o$ = 13 Gyr, consistent with observations.
Estimating this from the young stellar component at 2800 $\AA$ or from the
K-band luminosity density (with an assumed mass-to-light ratio) also give consistent results
for an assumed Salpeter mass-function.

\begin{figure}
\plottwo{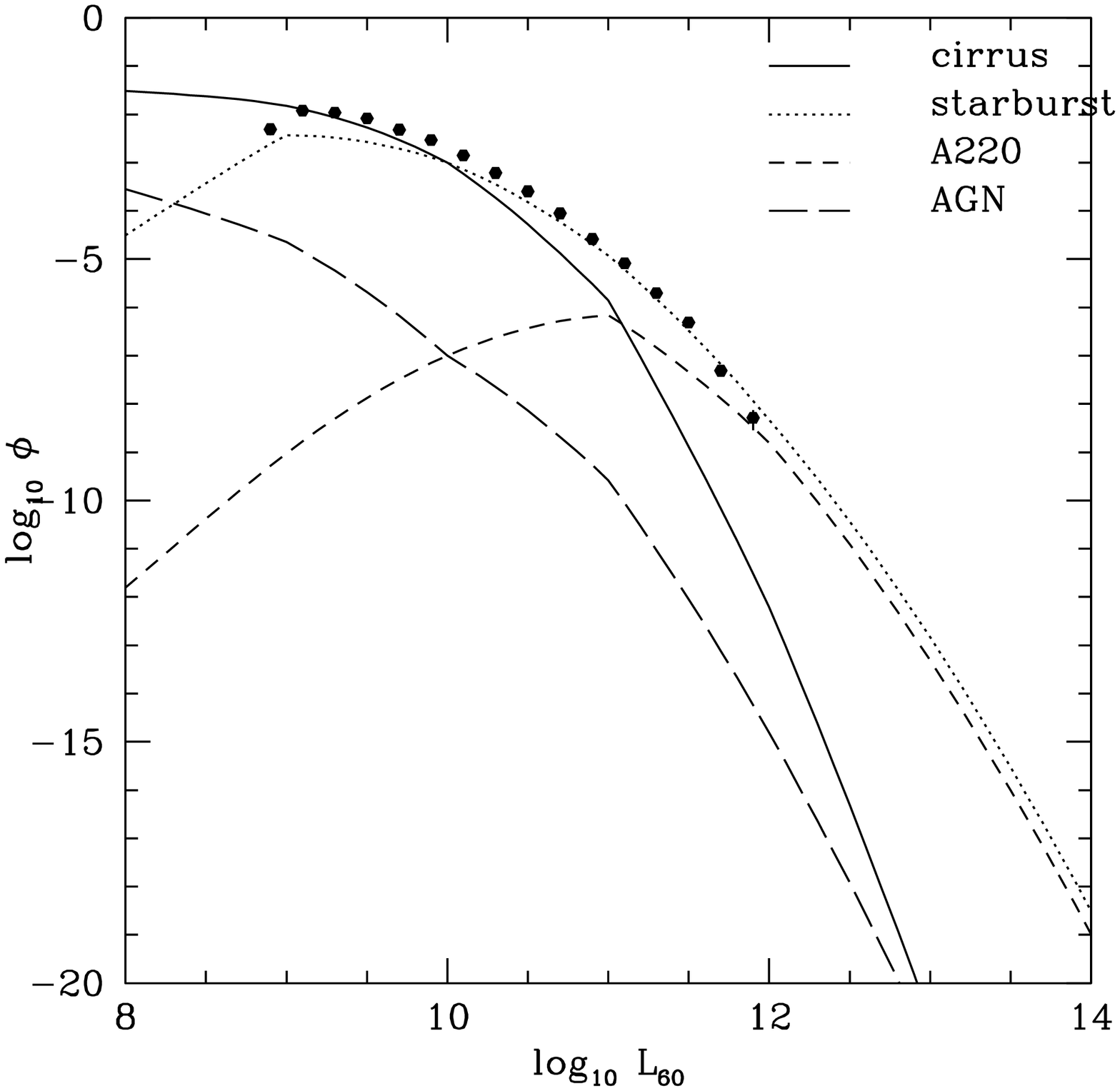}{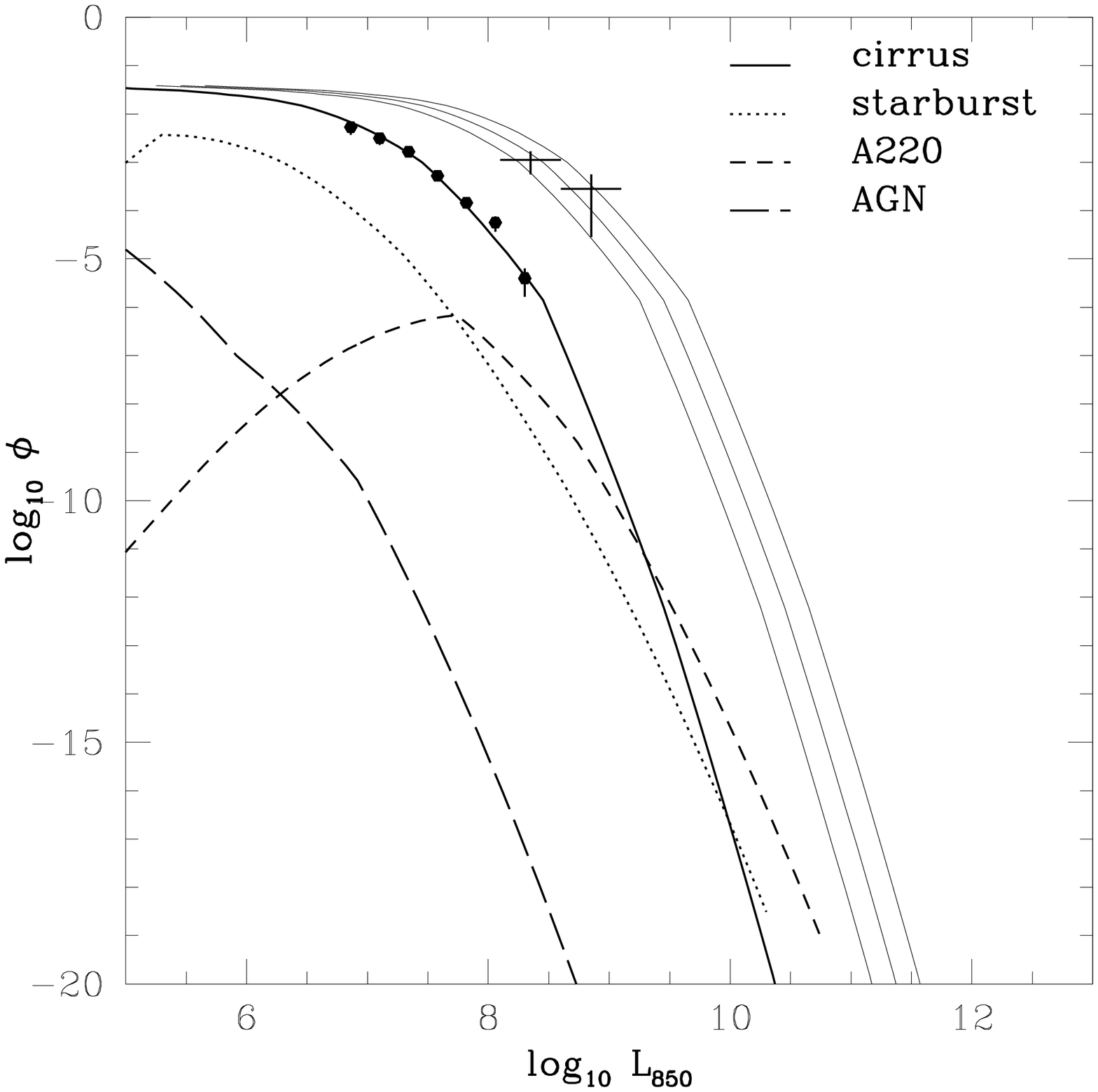}
\caption{
(L) Luminosity functions at 60 $\mu$m for the 4 spectral components.
Units of $\phi$ are $(Mpc)^{-3} dex^{-1}$, luminosity ($\nu L_{\nu}$) in solar units.
All luminosity functions are for $\Omega_o$ =1 model, $H_o$ = 100 $km s^{-1} Mpc^{-1}$.
Observed points are derived from PSCz data.  
(R) Luminosity functions at 850 $\mu$m for the 4 spectral components.  The filled circles are 
data from Dunne et al (2000).  The crosses are derived from the data of Hughes et al (1998)
for the HDF.}
\end{figure}

\begin{figure}
\plottwo{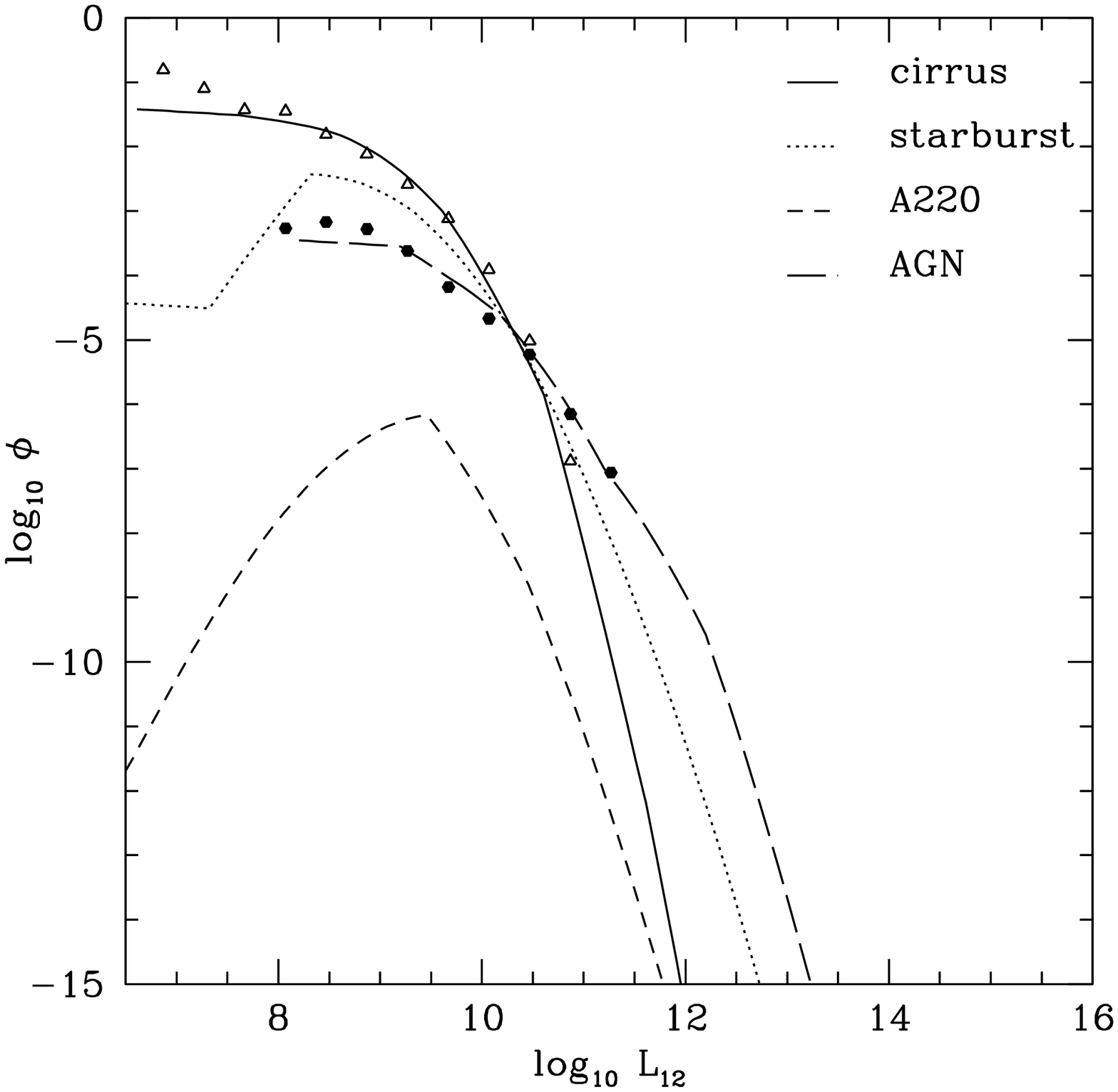}{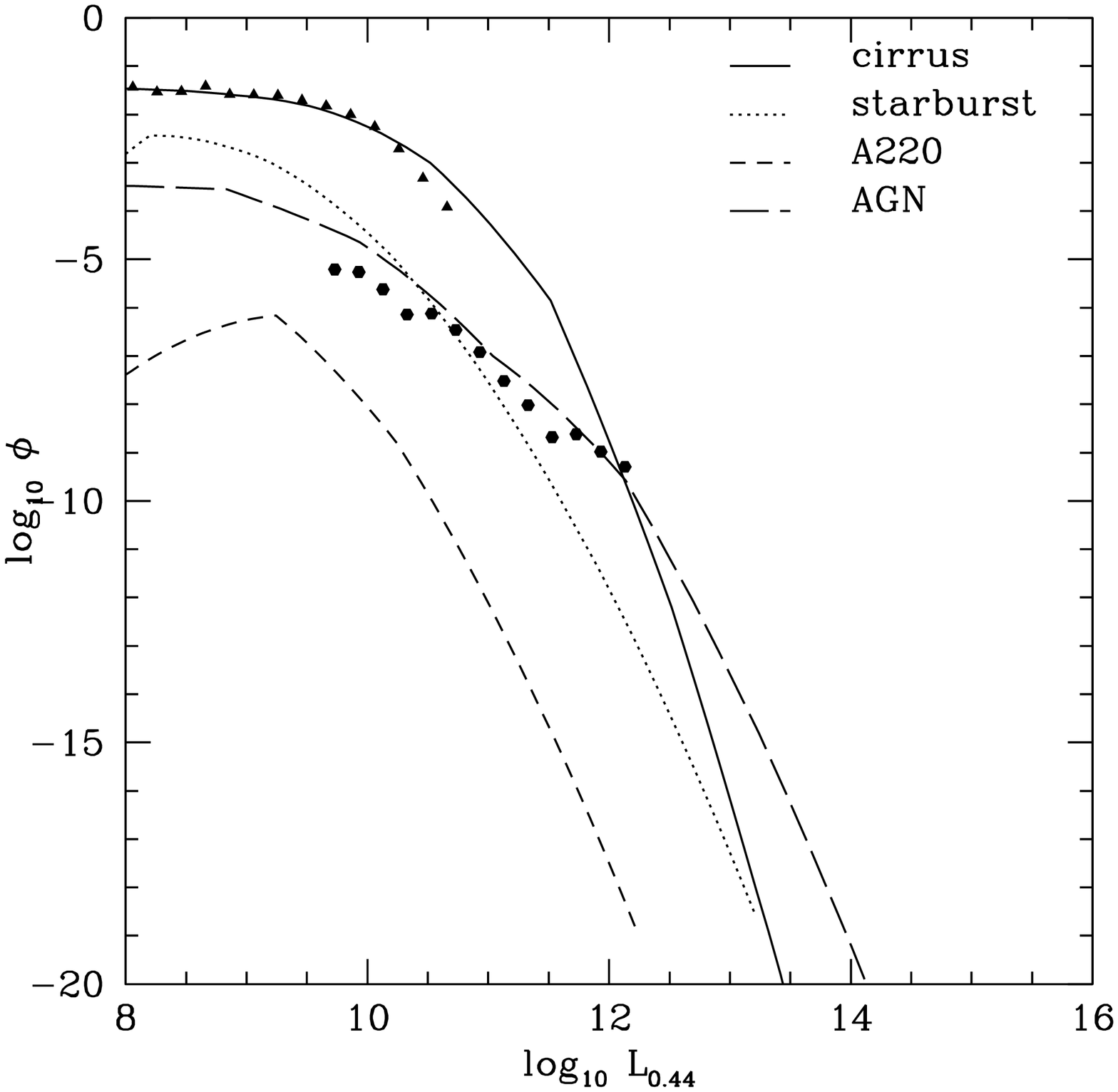}
\caption{
(L) Luminosity functions at 12 $\mu$m for the 4 spectral components.  Observed points
taken from Rush et al (1993) (filled circles: Seyferts, open triangles: non-Seyferts).
(R) Luminosity functions at 0.44 $\mu$m for the 4 spectral components.  Data for quasars derived
from PG sample and for galaxies from Loveday et al (1992).}
\end{figure}

\section{Predicted counts and integrated background spectrum from uv to submm}
  
Figure 5-8 shows the predicted source-counts in the 3 selected cosmological models, at
850, 175, 90, 60, 15, 2.2 and 0.44 $\mu$m.  The agreement with observations at infrared
and submillimetre wavelengths is extremely impressive.  Although the fits at 850 and 60
$\mu$m have been ensured by the least-squares procedure for determining P and Q, the
fits at 175, 90 and 15 $\mu$ are simply a consequence of the assumed seds and the
choice of the $t_i(L_{60})$.  There is not much difference
between the predictions of the 3 cosmological models at 60-850 $\mu$m.  At 15 $\mu$m there
is a difference between the models in the predicted numbers of sources at fluxes below 100 $\mu$Jy.  
The proportion of AGN dust tori at 12 $\mu$ agrees well that the data of Rush et al (1993) 
(15$\%$ brighter than 0.4 Jy).  Fig 7 shows that
the proportion of AGN at 15 $\mu$m is reasonably constant (15-20$\%$) for fluxes brighter than 3 mJy, but 
is predicted to fall rapidly towards fainter fluxes.  

The model does not at present have ingredients capable 
of accounting for the very faint K- and B-band galaxy counts.  This could be provided
either by a measure of density evolution (see below) or by steepening the faint-end luminosity function 
at z $>$ 1, which could in either case be attributed to a population of galaxies that had
merged into present-day galaxies.

\begin{figure}
\plottwo{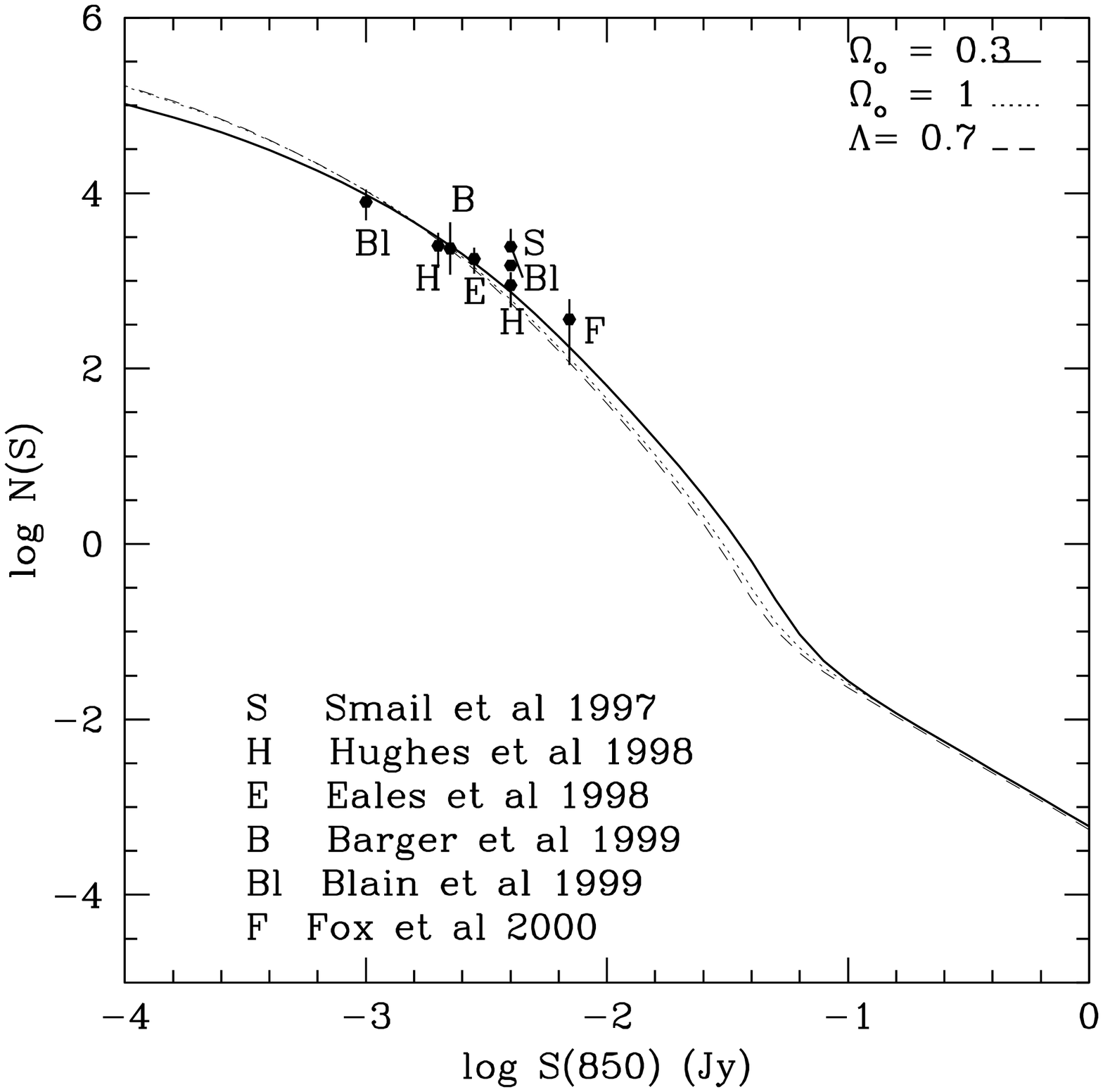}{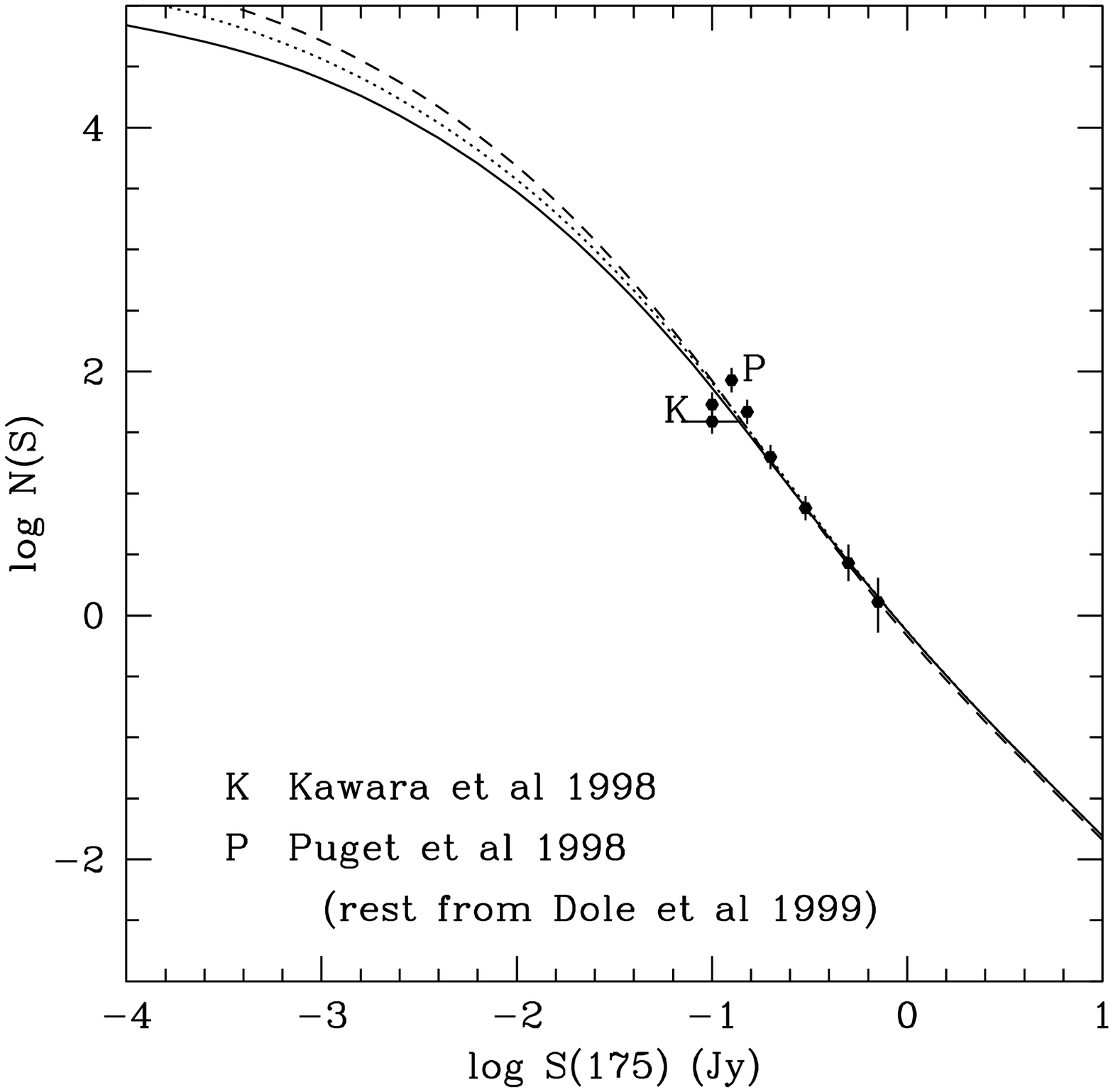}
\caption{
(L) Integral source counts at 850 $\mu$m.   Data are from Hughes et al (1998), Eales et al (1998),
 Smail et al (1997), Barger et al (1999), Blain et al (2000), Fox et al (2000).  
 The 3 models shown are, from bottom at faint fluxes, for $\Omega_o = 1$ and (P,Q) = (1.2, 5.4), 
 (solid curve), for $\Omega_o = 0.3$, (P,Q) = (2.1, 7.3) (dotted curve) and for $\Lambda$ = 0.7,
 (P,Q) = (3.0, 9.0) (broken curve).
(R) Source counts at 175 $\mu$m. Data points are from Kawara et al (1998), Guiderdoni et al (1998),
Dole et al (2000).}
\end{figure}

\begin{figure}
\plottwo{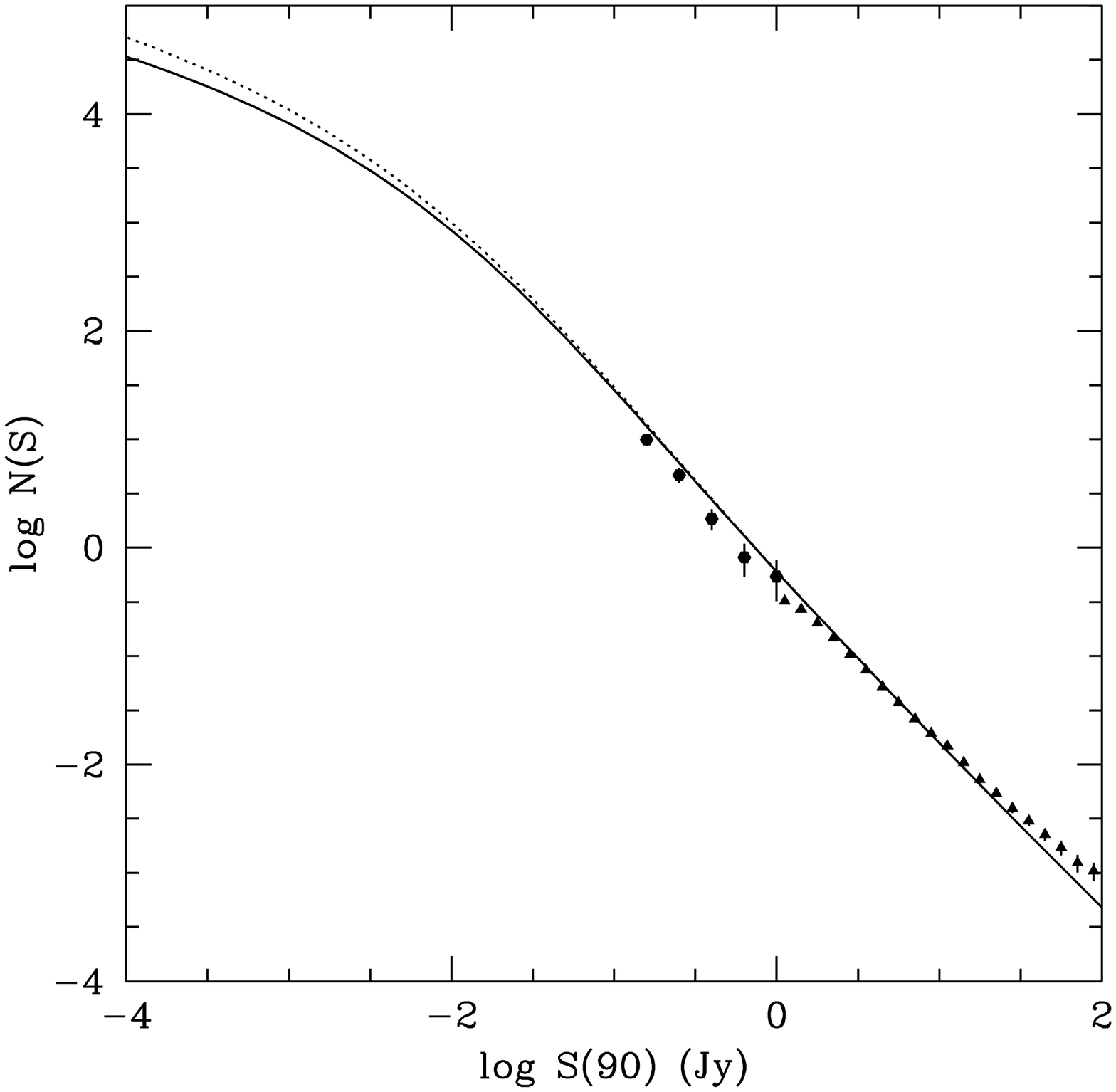}{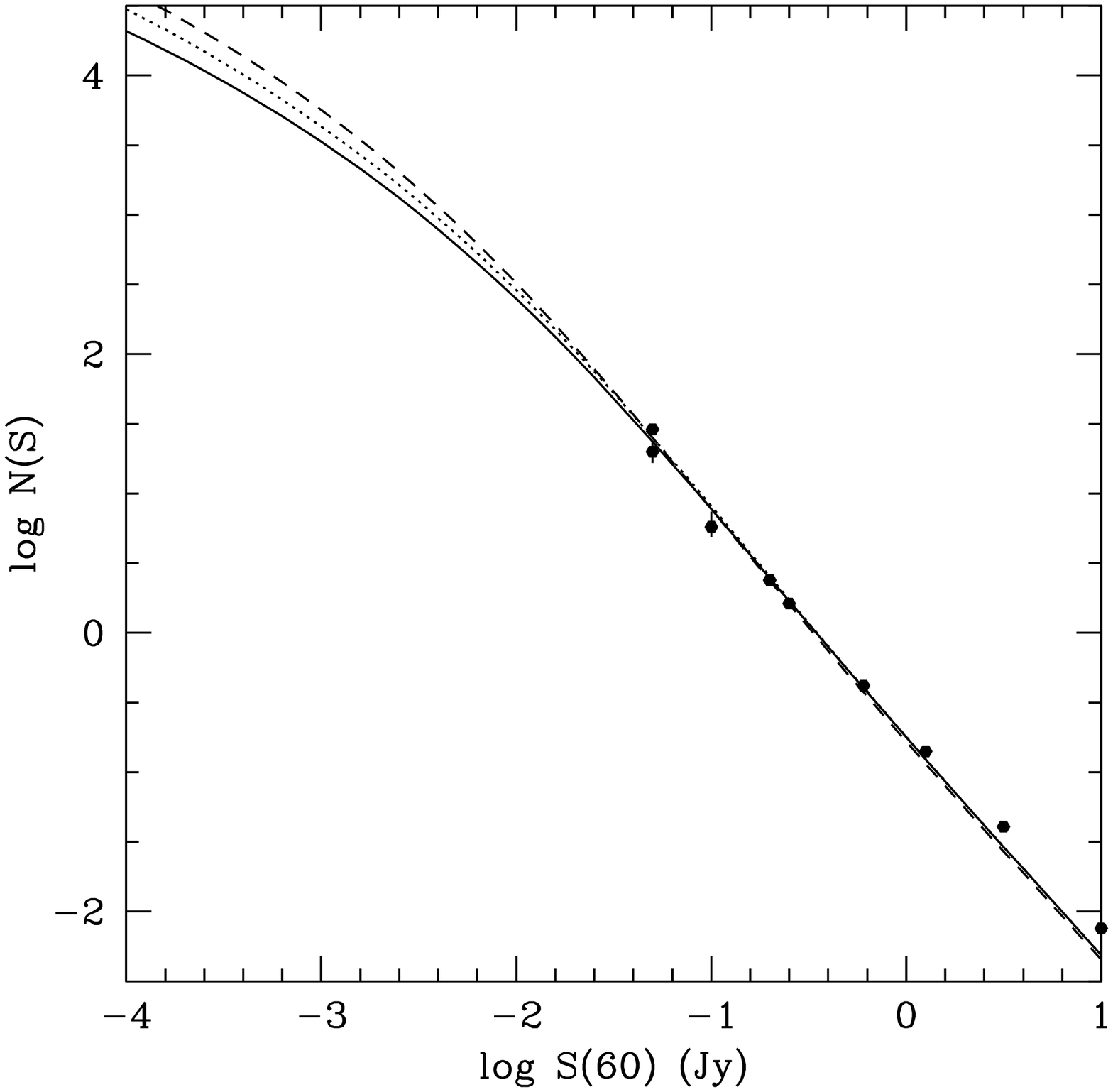}
\caption{
(L) Source-counts at 90 $\mu$m.  Data from IRAS PSCz (triangles) and ELAIS (filled circles)
are from Efstathiou et al (2000). (R) Source counts at 60 $\mu$m.   Data are from Lonsdale et al 
(1990) (at 0.2-10 Jy), Hacking and Houck 
(1987) (at 50-100 mJy), Rowan-Robinson et al (1991),
Gregorich et al (1995) (higher point at 50 mJy).  Models as in Fig 5.}
\end{figure}

\begin{figure}
\plotone{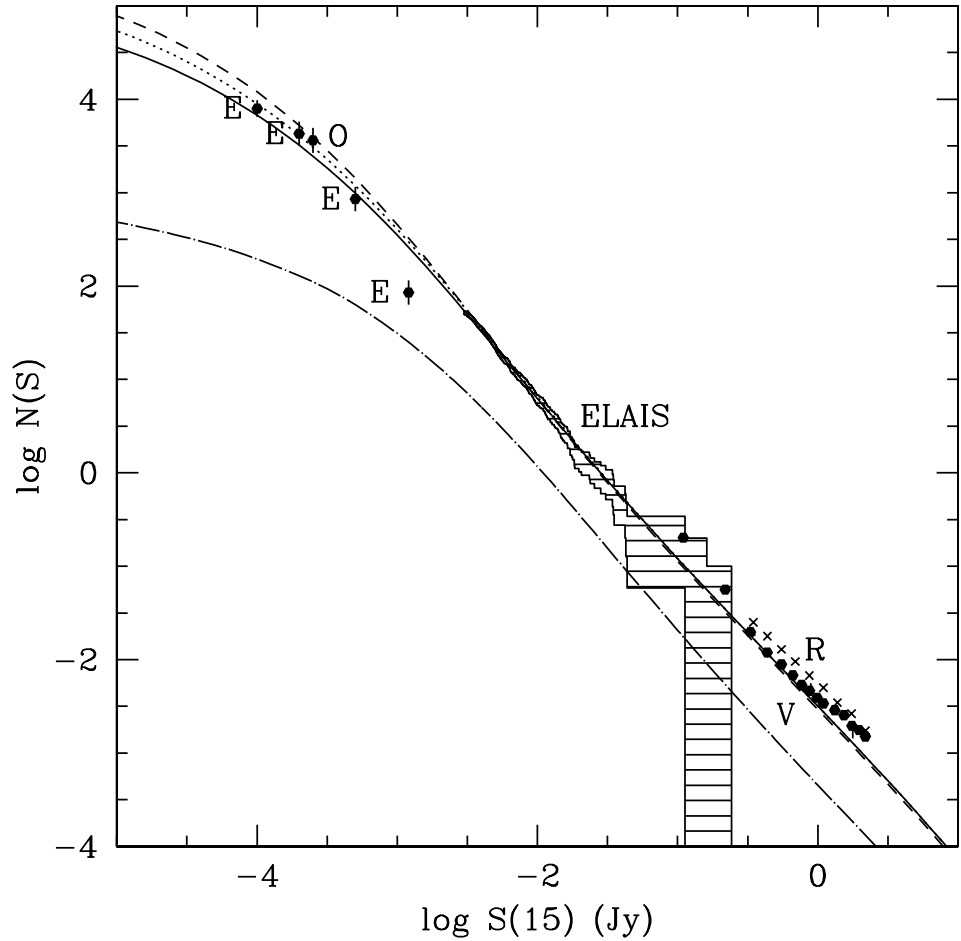}
\caption{
Source counts at 15 $\mu$m.  Data from Oliver et al (1997, O),
Serjeant et al (2000, ELAIS), Elbaz et al et al (2000, E) 
Rush et al (1993, R, crosses), Verma (2000, V, filled circles).
Models as in Fig 5.  The lower dash-dotted line shows counts of the AGN dust torus
population for the $\Omega_o$ = 1 model.}
\end{figure}

\begin{figure}
\plottwo{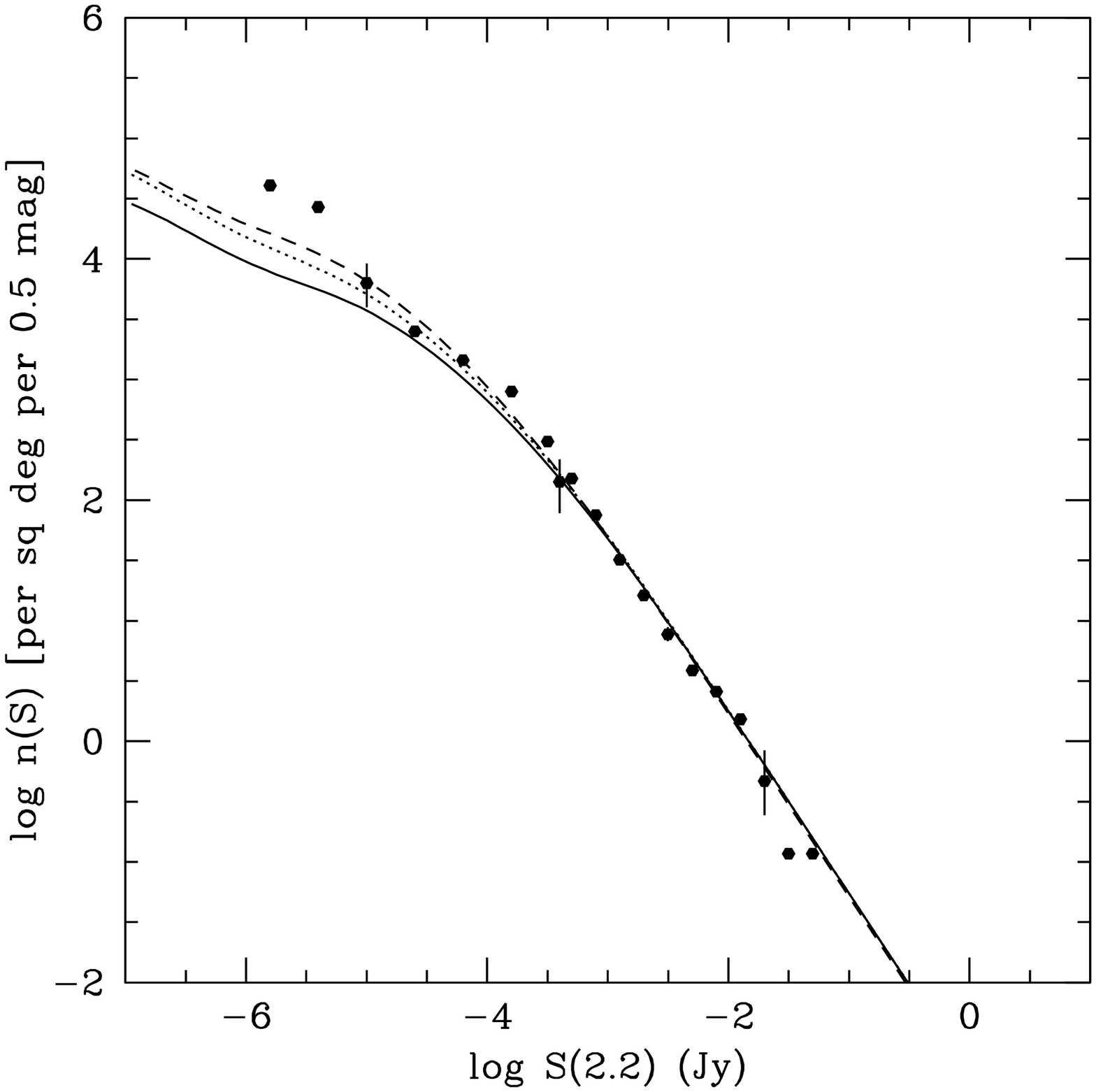}{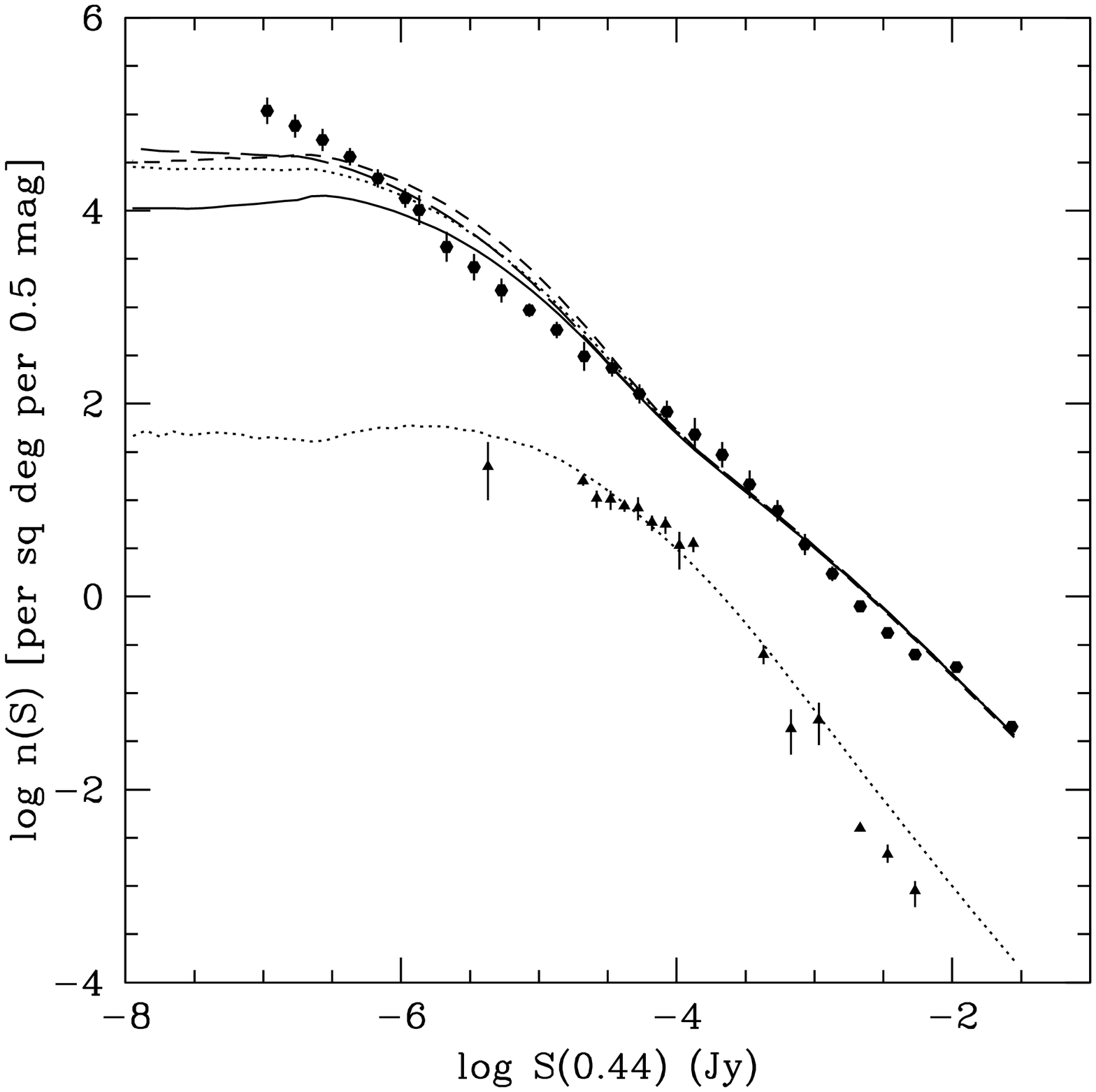}
\caption{
(L) Differential source counts at 2.2 $\mu$m.  Data from McCracken et al (2000).  
(R) Differential source counts at 0.44 $\mu$m.  Galaxy data (filled circles)
is from Metcalfe et al (1995), quasar data (triangles) from Boyle et al (1988).  
Models as in Fig 5.  The long-dashed line shows the effects of including density evolution 
(see text) on the
$\Omega_o$ = 1 model.  }
\end{figure}

Fig 9 show redshift distributions at selected wavelengths and fluxes 
%(complete output of the models is available at http://icstar5.ph.ac.uk/$\sim$mrr/countmodels).  
The median redshift at S(850 $\mu$m) = 2 mJy is predicted to be 2.2, significantly deeper than 
the prediction
for S(0.44 $\mu$m) = 0.1 $\mu$Jy (B = 26.6 m.).  This shows the power of the 850 $\mu$m surveys
and also the difficulty there will be in identifying the sources detected.

Figure 10 shows the predicted integrated background spectrum for the 3 cosmological
models.  All are consistent with the data, though the predictions of the $\Omega_o$ = 1
model are on the low side, while those of the $\Lambda$ = 0.7 model are on the high side.  
Figure 10 (R) shows, for the $\Omega_o$ = 1
model, the contribution of the different sed types to the background. The dominant contribution is
from the cirrus component at most wavelengths, so the prediction is that more of the
energy from starbursts is deposited in the general interstellar medium of a galaxy than
is absorbed in the early stages close to the location of the massive stars.  This dominance
by the cirrus component at submillimetre wavelengths also implies that many of the detected
sources should turn out to be rather extended (kiloparsec scales rather than the more
compact scales expected for nuclear starbursts).

\begin{figure}
\plottwo{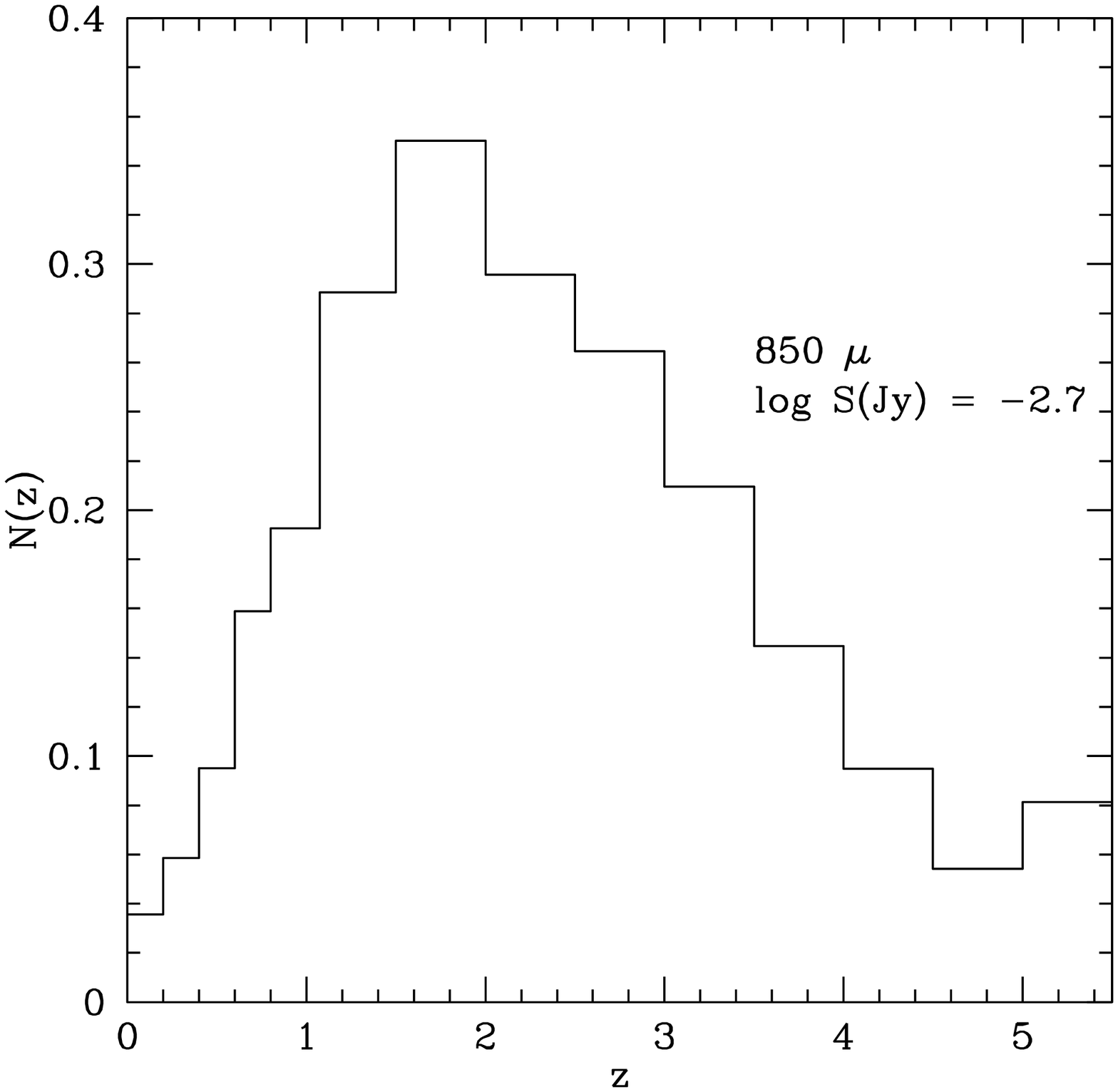}{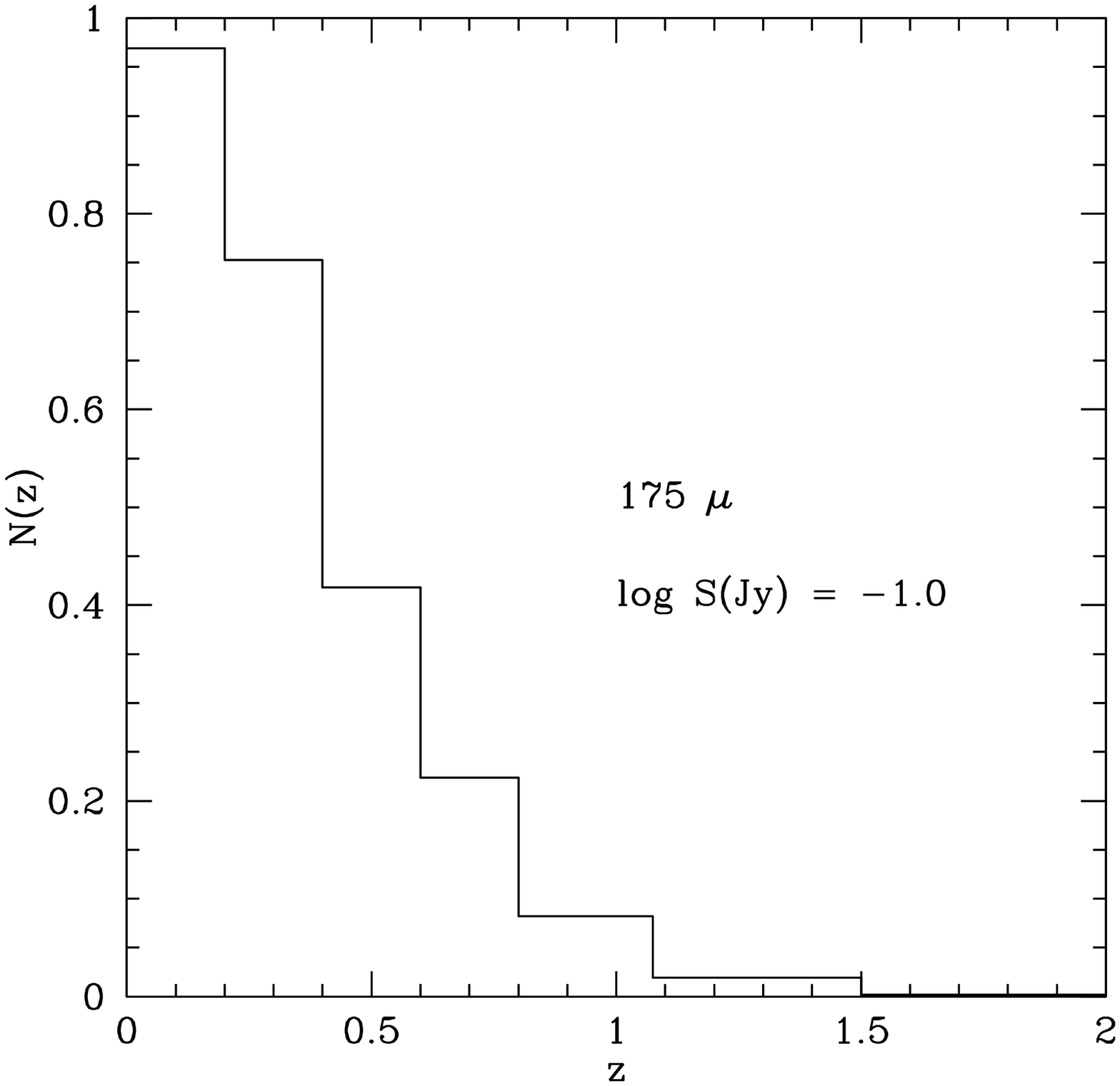}
\caption{
(L) Redshift distribution at 850 $\mu$m, $log_{10} S(Jy) = -2.7$. Bin centred at z = 5.25 
refers to z $>$ 5.  (R) Redshift distribution at 175 $\mu$m, $log_{10} S(Jy) = -1.0$. }
\end{figure}

\begin{figure}
\plottwo{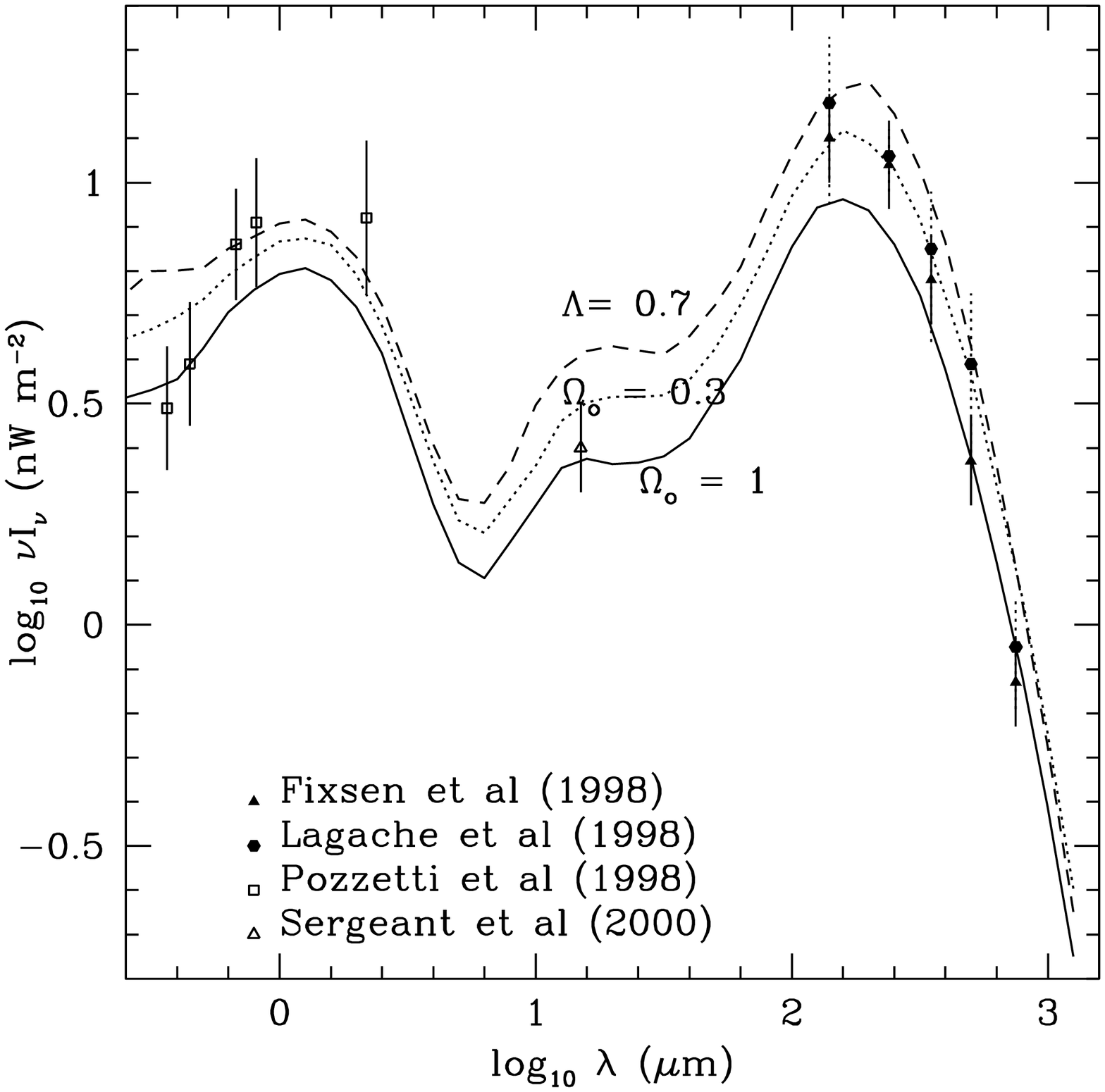}{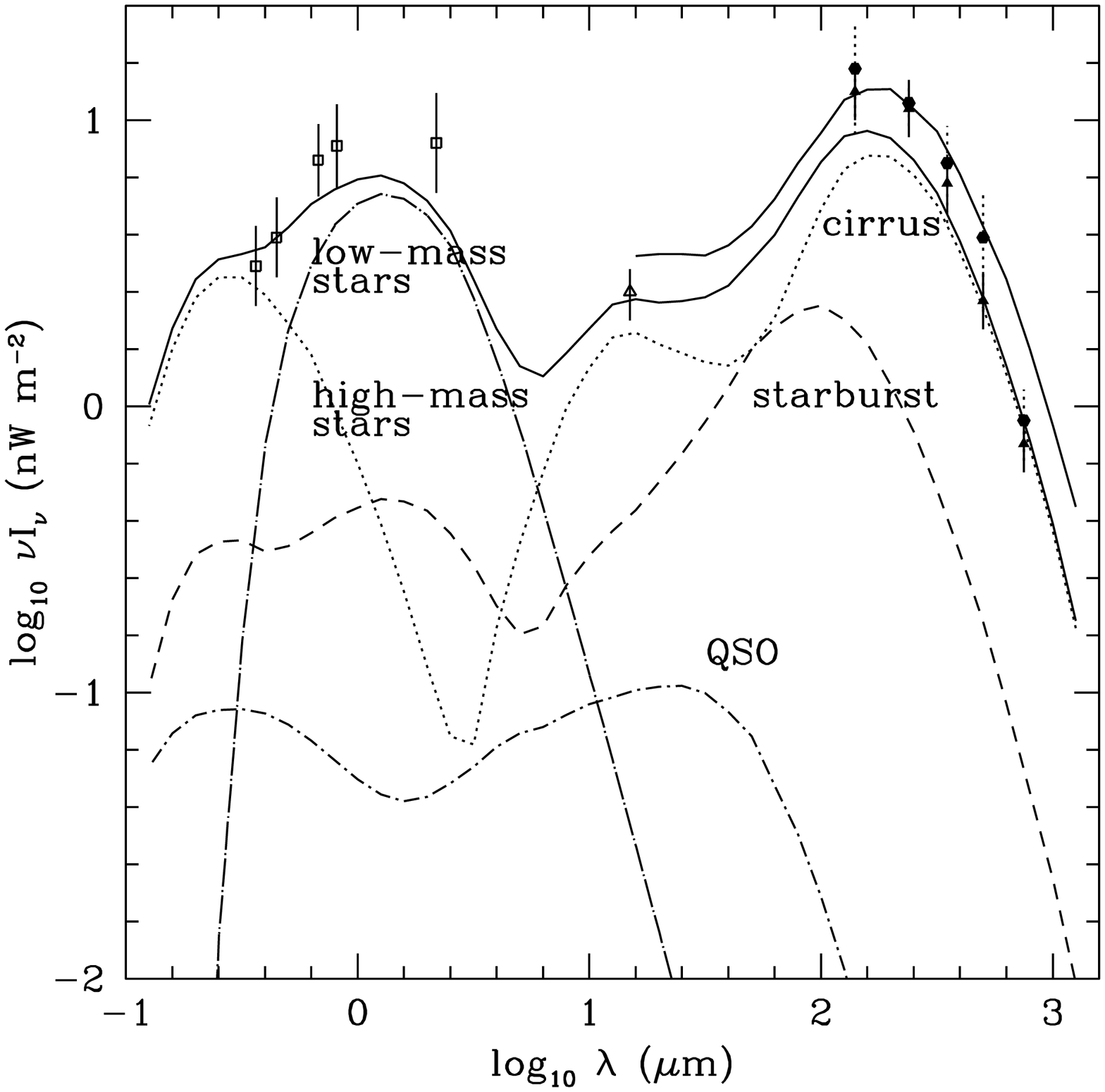}
\caption{
(L) Predicted spectrum of integrated background for same models as Fig 5.
(R) Predicted spectrum of integrated background for $\Omega_o = 1$ model,
showing contribution of the different components.   The contribution of
the Arp220-like starbursts is less than 0.01 $nW m^{-2}$ at all wavelengths.
The upper solid curve at long wavelengths shows the effect of including density evolution
(see text).
Data from Fixsen et al (1998)
(far ir, submm), Pozzetti et al (1998) (opt, uv).}
\end{figure}

\begin{figure}
\plotone{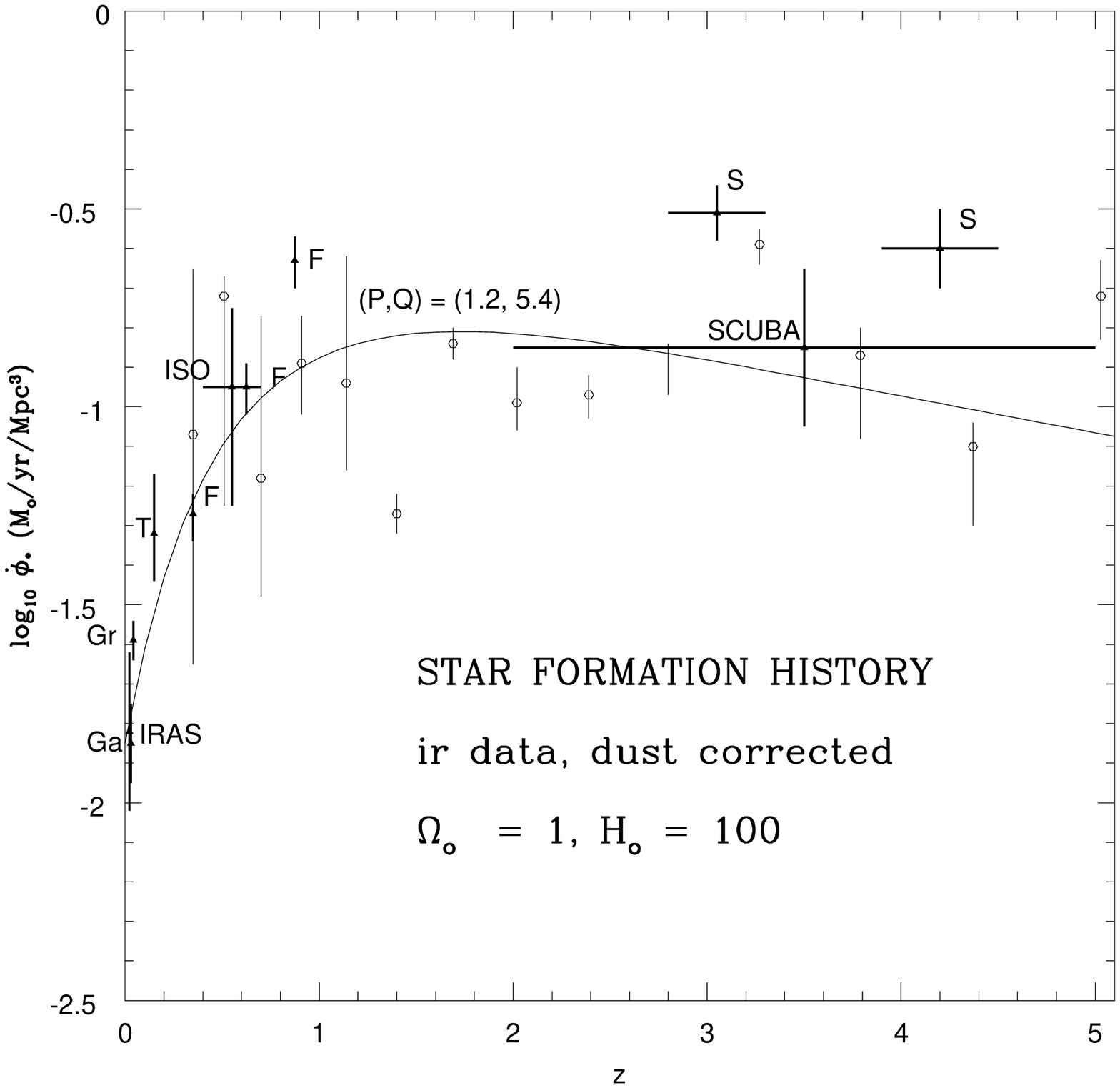}
\caption{Direct estimates of star formation rate, for $\Omega_o = 1$ case, from infrared, 
submm or radio data, or with
correction for the effects of insterstellar dust, as a function of z.  Data from Gallego et al (1995),
Rowan-Robinson (1997, revised: ISO), Gronwall (1998), Hughes et al (1998, revised: SCUBA), 
Treyer et al (1998), Flores et al (1999), Steidel et al (1999), and from photometric redshifts 
in the HDF (open circles, Rowan-Robinson 2000, in prep.).  The ISO and SCUBA HDF points have been
recalculated using a directly determined luminosity function (cf. Fig 3, right-hand panel). 
The solid curve is the best-fitting model to the counts and background spectrum.}
\end{figure}

\section{Effect of including density evolution}

In the framework of hierarchical, bottom-up galaxy formation scenarios, like those
based on cold dark matter, we expect that galaxies form as the result of mergers
between smaller fragments.  Thus we might expect to see a higher total density of
galaxies as we look back towards the past.  To test whether attributing part of the
star formation history to density evolution has a major effect on the predicted
counts and backgrounds, I have considered, for the $\Omega_o = 1 $ case,
a simple modification in which the
comoving density of galaxies varies with redshift as

$\rho(z) = \rho(0) (1+z)^n$.

For n=1, this means that the comoving number-density of galaxies at redshifts 2, 5 and 10 (our
assumed cutoff redshift), is increased by a factor 3, 6 and 11 compared with the present.
This would represent a very substantial degree of merging of galaxies over the observable 
range of redshifts.

Since the background intensity depends on the product of the luminosity and density 
evolution rates, the background spectrum will be unaltered if we simply increase P by $2n/3$.
At low redshifts the counts will be unaffected if P is increased by $n/3$.
I find that for a combined density evolution model with n = 1 and luminosity evolution 
with P = 1.5, Q = 5.4 (and $\Omega_o$ = 1) the fits to the counts at 15-850 $\mu$m
hardly change over the observed range,
but the predicted background is raised by about 0.2 dex, giving
better agreement with observations for this cosmological model.  For the $\Lambda$=0.7
model the fit to the background would be significantly worsened for models consistent 
with the observed counts.

\section{Discussion and Conclusions}

(1) I have developed a parameterized approach to the star formation history, which is sufficiently
versatile to reproduce many proposed model histories.  The model assumes that the evolution of 
the star formation rate manifests itself as pure luminosity evolution.  I have stressed the
importance of ensuring that the assumed luminosity function is consistent with available
60 $\mu$m redshift survey data and that the assumed spectral energy distributions are realistic.
The observed far ir and submm counts and
background then provide strong constraints on the model parameters.

The models consistent with infrared and submillimetre counts and backgrounds
tend to show a flat star formation rate from z = 1-3, consistent with the observed star
formation history derived from HDF galaxies and other data (Fig 11).
The most striking difference from previous modelling work is the dominant role of the
cirrus component at submillimetre wavelengths. 

Areas requiring further work are (i) the need to consider the evolution of the
shape of the seds, particularly for the cirrus component, with redshift.  The increased
star-formation rate at earlier times would tend to make the dust temperature higher, but this
is partially offset by the lower abundance of low-mass stars at earlier times. 
(ii) the AGN dust tori models require a further parameter, the orientation, and this may 
affect the AGN counts at optical wavelengths.
(iii) the tendency for the probability of finding an AGN component to increase with far
infrared luminosity is not fully reflected in the approach followed here. 

%\section*{Acknowledgements} 

\end{document}